\documentstyle[preprint,aps,epsfig]{revtex}
\tightenlines
\begin{document}
\draft
\preprint{\vbox{
\hbox{IFT-P.022/99}
\hbox{hep-ph/9903251}
\hbox{March 1999}
}}
\title{Spontaneous breaking of a global symmetry in a 331 model} 
\author{ J. C. Montero~\footnote{E-mail: montero@ift.unesp.br}, 
C. A. de S. Pires~\footnote{E-mail: cpires@ift.unesp.br} and 
V. Pleitez~\footnote{E-mail: vicente@ift.unesp.br}}  
\address{
Instituto de F\'\i sica Te\'orica\\
Universidade Estadual Paulista\\
Rua Pamplona 145\\ 
01405-900-- S\~ao Paulo, SP\\Brazil} 
\date{\today}
\maketitle
\begin{abstract}
In a 331 model in which the lepton masses arise from a scalar sextet 
it is possible to break spontaneously a global symmetry implying in a
pseudoscalar majoron-like Goldstone boson. 
This majoron does not mix with any other scalar fields and 
for this reason it does not couple, at the tree level, neither to the 
charged leptons nor to the quarks. Moreover, its interaction 
with neutrinos is diagonal. 
We also argue that there is a set of the 
pa\-ra\-me\-ters in which that the model can be consistent with the invisible 
$Z^0$-width and that heavy neutrinos can decay sufficiently rapid by
majoron emission having a lifetime shorter than the age of the universe.
\end{abstract}

\pacs{PACS numbers: 14.80.Mz; 12.60.-i; 12.60.Fr}

\section{Introduction}
\label{sec:intro}
In chiral electroweak models neutrinos can be massless at any order in 
perturbation theory if both conditions are supplied: 
no right-handed neutrinos are introduced and the total lepton number is 
conserved. If we do not assume the lepton number conservation we have two 
possibilities: we break it by hand, {\it i. e.}, explicit breaking or, we 
break it spontaneously. The later possibility implies the existence of a 
pseudoscalar Goldstone boson named majoron which  was firstly suggested in 
Ref.~\cite{cmp1} where a nonhermitian scalar singlet (singlet majoron model) 
was introduced, and in Ref.~\cite{gr} where a nonhermitian scalar triplet was 
introduced (triplet majoron model). 
Since the data of LEP~\cite{pdg} the triplet majoron model was ruled out.
The original model only considered one triplet and one doublet. 
In that model the majoron is a linear combination of doublet and triplet 
components but it is predominantly triplet. 
Hence, the lightest scalar ($R^0$) has a mass which is 
proportional to the VEV of the triplet and for this reason it is very small. 
Once we have the decay $Z^0\to R^0M^0$, where $M^0$ denotes the majoron, and 
since the only decay of the light scalar is $R^0\to M^0M^0$ there is an extra 
contribution the $Z^0$-invisible width. Its contribution
is exactly twice the contribution of a simple neutrino. Since the Higgs
scalars have only weak interaction they escape undetected. Hence, any 
experiment that counts the number of neutrino species by measuring the 
$Z^0$-invisible width automatically counts five neutrinos~\cite{ggn,concha}.

There are also possibilities involving only Higgs doublets, charged 
singlet scalars but they also need to include Dirac neutrino singlets. 
The minimal model of this 
sort was proposed in Ref.~\cite{bs} where an extra doublet scalar carrying 
lepton number was added (doublet majoron model). 
The new doublet does not couples to leptons.
The LEP data implies that at least a second doublet of the new type has
to be introduced~\cite{km}. In this sort of models since there are at least 
three doublets the lightest scalar $R^0$ can be assumed naturally to be 
heavier than the $Z^0$ avoiding the decay $Z^0\to R^0M^0$. 
In this majoron doublet model the lepton number violation take place at the 
same scale of the electroweak symmetry breaking.
It is also possible to consider a majoron model with one complex singlet, a
complex triplet and the usual $SU(2)$-doublet~\cite{sv1}. In this case 
the majoron can evade the LEP data since it can be mainly a singlet.

\section{The model}
\label{sec:m1}
Here we will consider a model with 
$SU(3)_C\otimes SU(3)_L\otimes U(1)_N$ symmetry with both exotic quarks and
only the known charged leptons~\cite{331}.
In this model in order to give mass to all fermions it is necessary to 
introduce three scalar triplets
\begin{mathletters}
\label{eqs1}
\begin{equation}
{\bf \chi }=\left( 
\begin{array}{c}
\chi ^{-} \\ 
\chi ^{--} \\ 
\chi ^0
\end{array}
\right) \sim \left( {\bf 3},{\bf -1}\right),\quad {\bf \rho }=\left( 
\begin{array}{c}
\rho ^{+} \\ 
\rho ^0 \\ 
\rho ^{++}
\end{array}
\right) \sim \left( {\bf 3},{\bf 1}\right), \quad
{\bf \eta }=\left( 
\begin{array}{c}
\eta ^0 \\ 
\eta _1^{-} \\ 
\eta _2^{+}
\end{array}
\right) \sim \left( {\bf 3},{\bf 0}\right) .
\label{e1}
\end{equation}
and a sextet
\begin{equation}
S=\left( 
\begin{array}{ccc}
\sigma _1^0 & \frac{h_2^{-}}{\sqrt2} & \frac{h_1^{+}}{\sqrt2} \\ 
\frac{h_2^{-}}{\sqrt2} & H_1^{--} & \frac{\sigma _2^0}{\sqrt2} \\ 
\frac{h_1^{+}}{\sqrt2} & \frac{\sigma _2^0}{\sqrt2} & H_2^{++}
\end{array}
\right) \sim \left( {\bf 6},{\bf 0}\right).
\label{s}
\end{equation}
\end{mathletters}

Although we can assign a lepton number to the several scalar fields we 
prefer to use  the global quantum number ${\cal F}=L+B$~\cite{pt}.
It is clear that the model needs only one global quantum number and not four
as in the standard model {\it i.e.}, family lepton number $L_i,\; (i=e,\mu,
\tau$ with $L=\sum_iL_i$) and the baryonic number $B$. The quantum number 
${\cal F}$ coincide with $L$ and $B$ for the known particles but implies the 
assignation of a single quantum number to the other particles.

The most general gauge and ${\cal F}$ conserving scalar potential is 
\begin{eqnarray}
V&=&\mu^2_1\eta^\dagger\eta+\mu^2_2\rho^\dagger\rho+\mu^2_3\chi^\dagger\chi+
\mu^2_4{\mbox Tr}(S^\dagger S)+
\lambda_1(\eta^\dagger\eta)^2+\lambda_2(\rho^\dagger\rho)^2+
\lambda_3(\chi^\dagger\chi)^2 \nonumber \\ &+& 
\eta^\dagger\eta[\lambda_4\rho^\dagger\rho+\lambda_5\chi^\dagger\chi]+
\lambda_6(\rho^\dagger\rho)(\chi^\dagger\chi)+
\lambda_7(\eta^\dagger\rho)(\rho^\dagger\eta)+
\lambda_8(\eta^\dagger\chi)(\chi^\dagger\eta)+
\lambda_9(\rho^\dagger\chi)(\chi^\dagger\rho)\nonumber \\ &+&
\lambda_{10}({\mbox Tr}S^\dagger S)^2+\lambda_{11}{\mbox Tr}[(S^\dagger S)^2]
+{\mbox Tr}(S^\dagger S)[\lambda_{12}(\eta^\dagger\eta)+
\lambda_{13}(\chi^\dagger\chi)+\lambda_{14}(\rho^\dagger\rho)]
\nonumber \\ &+&
[\lambda_{15}\epsilon^{ijk}\chi_i(S\chi^\dagger)_j\eta_k+
\lambda_{16}\epsilon^{ijk}\rho_i(S\rho^\dagger)_j\eta_k+
\lambda_{17}\epsilon^{ijk}\epsilon^{lmn}S_{il}S_{jm}\eta_n\eta_k+H.c.]
\nonumber \\ &+&
\lambda_{18}\chi^\dagger SS^\dagger\chi+\lambda_{19}\eta^\dagger S
S^\dagger
\eta+\lambda_{20}\rho^\dagger SS^\dagger \rho+
\left[\frac{f_1}{2}\epsilon^{ijk}\eta_i\rho_j\chi_k+
\frac{f_2}{2}\chi^TS^\dagger\rho
+H.c.\right].
\label{potential}
\end{eqnarray}

Terms like the quartic $(\chi^\dagger\eta)(\rho^\dagger\eta)$, 
$\chi S\eta^\dagger\rho$, $\eta S\eta^\dagger\eta, \chi\rho SS$ and the 
trilinear $\eta S^\dagger \eta, SSS$ do not conserve the ${\cal F}$ 
quantum number (or the lepton $L$) and they will not be considered here. 

The minimum of the potential must be studied after the shifting of the
neutral components of the three scalar multiplets. Hence, we
redefine the neutral components in Eqs.~(\ref{eqs1}) as follows:
\begin{equation}
\eta ^0\to \frac{1}{\sqrt{2}}\left( v_\eta +R^0_1 +iI^0_2\right) ,\;\;
\rho ^0\to \frac{1}{\sqrt{2}}\left(v_\rho  +R^0_2 +iI^0_2\right), 
\;\;
\chi^0\to \frac{ 1}{\sqrt{2}}\,
\left( v_\chi+R^0_3+iI^0_3\right),
\label{vev}
\end{equation}
and
\begin{equation}
\sigma^0_1\to \frac{1 }{\sqrt{2}}\,
\left(v_{\sigma_1} +R^0_4+iI^0_4\right),
\quad
\sigma^0_2\to \frac{1 }{\sqrt{2}}\,
\left(v_{\sigma_2} +R^0_5+iI^0_5\right),
\label{vevs1}
\end{equation}
where $v_a$ (with $a=\eta,\rho,\chi,\sigma_1,\sigma_2$) are considered real 
parameters for the sake of simplicity.
The ${\cal F}$ number attribution is the following:
\begin{eqnarray}
{\cal F}(U^{--})&=&{\cal F}(V^{-}) = - {\cal F}(J_1)= {\cal F}(J_{2,3})=
{\cal F}(\rho^{--}) = {\cal F}(\chi^{--}) ={\cal F}(\chi^{--}) \nonumber \\
&=& {\cal F}(\sigma_1^0)={\cal F}(h^-_2)={\cal F}(H^{--}_1)={\cal F}(H^{--}_2)=2,
\label{efe}
\end{eqnarray}
where $J_1$ ($J_{2,3}$) are exotic quarks of charge 5/3 (-4/3) present in
the model and we have included them by completion. For leptons and the known 
quarks ${\cal F}$ coincides with the total lepton and baryon numbers, 
respectively. All the other fields have ${\cal F}=0$. 
As we said before, all terms in Eq.~(\ref{potential}) conserve the ${\cal F}$ 
quantum number.
However, if we assume that  $\langle \sigma^0_1\rangle\not=0$
we have the spontaneous breakdown of ${\cal F}$ and the corresponding
pseudoscalar, the majoron $M^0$, as we will show below~\cite{tj}.

In a model with several complex scalar fields, as is the case of the 331 
model~\cite{331}, if the lepton number is spontaneously broken one of the
neutral scalars is a Goldstone boson associated with the global symmetry 
breaking. With respect to the $SU(2)_L\otimes U(1)_Y$ gauge symmetry, this 
model has naturally three doublets: $(\rho^+,\rho^0)^T,(\eta^0,\eta^-)^T$, and 
$(h^+,\sigma_2^0)^T$; one triplet: 
\begin{equation}
\left(\begin{array}{cc}
\sigma^0_1 & \frac{h^-_2}{\sqrt2}\\
\frac{h^-_2}{\sqrt2} &H^{--}_1
\end{array}
\right);
\label{tsu2}
\end{equation}
and two singlets: $\chi^{--}_2$ and $\chi^0$.

The mass matrix in the real sector in the basis 
$(R^0_1,R^0_2,R^0_3,R^0_4,R^0_5)^T$, is given by the symmetric matrix
\begin{eqnarray}
&&m_{11}=\lambda_1v^2_\eta-\frac{\lambda_{16}}{4\sqrt2 }
\frac{v^2_\rho v_{\sigma_2}}{v_\eta}+\frac{1}{4\sqrt{2}v_\eta}
(\lambda_{15}v_\chi v_{\sigma_2}-f_1v_\rho )v_\chi+\frac{t_\eta}{2v_\eta},
\nonumber \\ &&
m_{22}=\lambda_2v^2_\rho-\frac{1}{8\sqrt2 v_\rho}(2f_1 v_\eta +
f_2v_{\sigma_2})v_\chi+\frac{t_\rho}{2v_\rho},
\nonumber \\ &&
m_{33}=\lambda_3v^2_\chi-\frac{1}{8\sqrt2 v_\chi}(2f_1v_\eta +
f_2v_{\sigma_1})v_\rho+\frac{t_\chi}{2v_\chi},\quad
m_{44}=(\lambda_{10}+\lambda_{11})v^2_{\sigma_1}+
\frac{t_{\sigma_1}}{2v_{\sigma_1}},
\nonumber \\ &&
m_{55}=(\lambda_{10}+\frac{\lambda_{11}}{2})v^2_{\sigma_2}
-\frac{1}{4\sqrt2 v_{\sigma_2}}\lambda_{16} v^2_\rho v_\eta 
-\frac{1}{8v_{\sigma_2}}(f_2v_\rho+\sqrt2
\lambda_{15}v_\eta)v_\chi+\frac{t_{\sigma_2}}{2v_{\sigma_2}}, 
\nonumber \\ &&
m_{12}=\frac{1}{2}(\lambda_4v_\eta +\frac{\lambda_{16}}{\sqrt2}v_{\sigma_2})
v_\rho+\frac{f_1}{4\sqrt2}v_\chi, 
\nonumber \\ &&
m_{13}=\frac{1}{2}(\lambda_5 v_\eta -
\frac{\lambda_{15}}{\sqrt2} v_{\sigma_2})v_\chi+\frac{f_1}{4\sqrt2}v_\rho,
\nonumber \\ &&
m_{14}=(\lambda_{12}+\lambda_{19})\frac{v_\eta v_{\sigma_1}}{2},
\nonumber \\ &&
m_{15}=\frac{1}{2}(\lambda_{12} -2\lambda_{17})v_\eta v_{\sigma_2}-
\frac{\lambda_{15}}{4\sqrt2}v^2_\chi+\frac{\lambda_{16}}{4\sqrt2}v^2_\rho,
\nonumber \\ &&
m_{23}=\frac{\lambda_6}{2}v_\rho v_\chi+\frac{f_1}{4\sqrt2}v_\eta+
\frac{f_2}{8}v_{\sigma_2},
\nonumber \\ &&
m_{24}=\frac{\lambda_{14}}{2}v_\rho v_{\sigma_1},
\nonumber \\ &&
m_{25}=(2\lambda_{14}+\lambda_{20})\frac{v_\rho v_{\sigma_2}}{4}+
\frac{\lambda_{16}}{2\sqrt2}v_\eta v_\rho+\frac{f_2}{8}v_\chi, \quad
m_{34}= \frac{\lambda_{13}}{2}v_\chi v_{\sigma_1}, 
\nonumber \\ &&
m_{35}= \frac{\lambda_{13}}{2}v_\rho v_{\sigma_2} -
\frac{1}{4\sqrt2}(\lambda_{15} v_\eta- \lambda_{18}
v_{\sigma_2})v_\chi+\frac{f_2}{8}v_\rho,
\nonumber \\ &&
m_{45}= \lambda_{10}v_{\sigma_1} v_{\sigma_2}. 
\label{mmr}
\end{eqnarray}
The tadpole equations $t_a$ where $a=\eta,\rho,\chi,\sigma_1,\sigma_2$ are  
given in the Appendix. The conditions for a extreme of the potential are 
$t_a=0$. 
Assuming that the matrix $m_{ij}$ above is diagonalized by an orthogonal 
matrix ${\cal O}$, the relation among symmetry ($R^0_i$) and mass ($H^0_j$) 
eigenstates is $R^0_i={\cal O}_{ij}H^0_j$; $i,j=1,2,3,4,5$. The masses 
$m_{H_j}$ can vary, depending on a 
fine tuning of the parameters, from a few GeVs till 2 or 3 TeV (typical 
values of the energy scale at which the break down of the $SU(3)_L$ symmetry
does occur). Also the arbitrary orthogonal matrix ${\cal O}$ is not 
necessarily almost diagonal. Denoting the lightest Higgs as $H^0_1$ two
extreme possibilities are compatible with the LEP data: $R_4=
({\cal O}^{-1})_{41}H^0_1+\cdots,$ with 
$({\cal O}^{-1})_{41}\ll1$, 
if $m_{H_1}<M_Z$; or $R_4\approx H^0_1+\cdots$, that is 
$({\cal O}^{-1})_{41}\approx1$, if $M_{H_1}>M_Z$. Intermediate values for the 
mass $m_{H_1}$ and the mixing angles have been ruled out by the LEP data 
(see below). 

The symmetric mass matrix of the imaginary part in the basis 
$(I^0_1,I^0_2,I^0_3,I^0_4,I^0_5)^T$ reads
\begin{eqnarray}
&&M_{11}= -\frac{\lambda_{16}}{4\sqrt{2}}\frac{v^2_\rho v_{\sigma_2}}
{v_\eta} +\lambda_{17}v^2_{\sigma_2}
+ \frac{1}{4\sqrt{2}}(\lambda_{15}v_\chi v_\eta-
f_1v_\rho) \frac{v_\chi}{v_\eta}+\frac{t_\eta}{2v_\eta}, 
\nonumber \\&&
M_{22}=-\frac{1}{8}(\sqrt2 f_1 v_\eta +f_2v_{\sigma_2}\frac{v_\chi}{v_\rho})
+\frac{t_\rho}{2v_\rho},\quad
M_{33}=-\frac{1}{8}(\sqrt2 f_1 v_\eta+f_2 v_{\sigma_2})\frac{v_\rho}{v_\chi}
+\frac{t_\chi}{2v_\chi},
\nonumber \\ &&
M_{44}=\frac{t_{\sigma_1}}{2v_{\sigma_1}},\quad
M_{55}=\frac{1}{4\sqrt2}(\lambda_{15}v^2_\chi-\lambda_{16}v^2_\rho)
\frac{v_\eta}{v_{\sigma_2}}+\lambda_{17}v^2_\eta-\frac{f_2}{8}\frac{v_\rho
v_\chi}{v_{\sigma_2}}+\frac{t_{\sigma_2}}{2v_{\sigma_2}},
\nonumber \\&&
M_{12}=-\frac{f_1}{4\sqrt2}v_\chi,\quad M_{13}=
-\frac{f_1}{4\sqrt2}v_\rho,\quad M_{14}=0,\quad
M_{15}=\frac{\lambda_{15}}{4\sqrt{2}} v_\eta-\frac{f_2}{8}v_{\sigma_2},
\nonumber \\&&
M_{23}=-\frac{1}{8}(\sqrt{2}f_1v_\eta+f_2v_{\sigma_2}),
\quad M_{24}=0,\quad M_{25}=\frac{f_2}{8}v_\chi,\quad
M_{34}=0,\quad M_{35}=\frac{f_2}{8}v_\rho, \nonumber \\ && M_{45}=0.
\label{mmi}
\end{eqnarray}

In Eqs.~(\ref{mmr}) and (\ref{mmi}) when all $v_a\not=0,\;a=\eta,
\rho,\chi,\sigma_1,\sigma_2$ then we can use $t_a=0$. 
In Eqs.~(\ref{mmi}) there are three Goldstone boson. 
Notice however that since  $M_{4i}=0$,
the component $I^0_4$ has a zero mass, {\it i.e.}, it is 
an extra Goldstone boson which decouples in the sense that it does not mix 
with the other $CP$-odd scalars. Hence, $I^0_4$ is the majoron field. 
Hereafter it will be denoted $M^0$. 
The submatrix $4\times 4$ has still two other Goldstone bosons which are
related to the masses of $Z^0$ and $Z^{'0}$. 
Hence, although the majoron in the present model is a triplet under the 
subgroup $SU(2)$ it does not mix with the other imaginary fields. 

Hence, as in the singlet majoron model ours has no couplings with 
fermions (charged leptons and quarks). Moreover, as we said before, the real 
component can be heavier than the $Z^0$. It is easy to understand this. 
If $v_{\sigma^0_1}=0$, the tadpole equation in Eq.~(A4) must be 
replaced in the mass matrices in Eqs.~(\ref{mmr}) and (\ref{mmi}). 
In this case the $\sigma^0_1$ fields consists of two mass-degenerate fields 
$R^0_4$ and $I^0_4$ with mass
\begin{equation}
m^2_{R_4}=m^2_{I_4}=
\lambda_{10}v^2_{\sigma_2}+ \frac{1}{2}(\lambda_{12}+\lambda_{19})v^2_\eta
+\frac{\lambda_{13}}{4}v^2_\chi+\frac{\lambda_{18}}{2}v^2_\rho.
 \label{mam}
\end{equation}
The mass in Eq.~(\ref{mam}) can be large because it depends on $v^2_\chi$.
When $ v_{\sigma_1}\not=0$ is used, the degeneration in mass of 
$R^0_4$ and $I^0_4$ is broken,
the imaginary part becomes the majoron and the real part has a mixing with the
other real neutral components, which include several fields 
transforming under $SU(2)_L\otimes U(1)_Y$ as doublets, 
($\eta^0,\rho^0,\sigma^0_2$), and one singlet, ($\chi^0$). This also happens 
in the one-singlet--one-doublet--one-triplet model when the triplet does not 
gain a VEV~\cite{diaz}. 
Notice that unlike $v_{\sigma_1}$, if $v_{\sigma_2}=0$ the condition in 
Eq.~(A5) forces $f_2=0$. All the other VEVs has to be nonzero
in order to have a consistent breaking of the $SU(3)$ symmetry.

\section{phenomenological consequences and conclusions}
\label{sec:feno}
In the present model the interaction of the majoron with the $Z^0$ 
( which is of the form $Z^0M^0H^0_j$), is given by
\begin{equation}
 {\cal O}_{4j}\,\frac{(\sqrt{2}G_F)^{\frac{1}{2}}}{c_W}\,
M_W(p_{M^0}-p_j)^\mu,
\label{zmi}
\end{equation}
where $p_{M^0}$ and $p_j$ are the momenta of the majoron and the physical
real scalars $H^0_j$, respectively. We see that if it is allowed, the 
contribution of the decay mode $Z^0\to H^0_1M^0$, where $H^0_1$ is the 
lightest Higgs scalar, is $2\vert{\cal O}_{41}
\vert^2$ times that of the neutrino antineutrino. Hence, as we said 
before, the value of the 
mixing matrix element ${\cal O}_{4j}$ is constrained appropriately: 
$2{\cal O}_{4j}^2\sim10^{-4}$, to make the model consistent with the LEP
data {\it i.e.}, now 
$\Gamma_Z\to H^0_1M^0$ (where $H^0_1$ is assumed the lightest scalar) could be
reduced to an acceptable 
level. More interesting, however, is the fact that in this model 
the $Z^0\to H^0_1M^0$ might be kinematically forbidden since $H_1$ can be 
heavier than the $Z^0$ as it was discussed above.

It is well known that if neutrinos are massive particles the 
thermal history of the universe strongly constrains the mass of stable 
neutrinos, {\it i.e.}, $m_\nu<100$ eV for light neutrinos or a few GeV for 
heavy ones~\cite{eu}. 
One of the ways in which the cosmological constraints on neutrino masses 
can be altered is when the lepton number is broken globally given rise to the 
majoron field: heavy neutrinos can decay sufficiently rapid by 
majoron emission, thereby given negligible contributions to the mass density 
of the universe~\cite{cmp2}.
Let us denote $\nu_h$ ($\nu_l$) heavy (light) neutrinos and look for  
$\nu\to \nu'+M^0$  decays in the present model. Those decays, as in the 
triplet majoron model, are completely forbidden at the tree 
level too (there is neither $\nu\to \nu'+\gamma$ nor $\nu\to 3\nu'$ decays 
at the lowest order). 

Here we will denote, as usual, $W^+$ the vector boson which coincides with the 
respective boson of the standard model, {\it i.e.}, it couples to the usual 
charged current in the lepton and the quark sectors and also satisfies 
$M^2_W/M^2_Z=c^2_W$; and $V^+$ denotes the vector boson which couples charged 
leptons with anti-neutrinos or the known quarks with the exotic ones.
If the lepton number is not spontaneously broken $W^+$ and $V^+$ do not 
couples to one another. However, a mixing between both $W^+$ and $V^+$ arises 
when the lepton number is spontaneously broken. 
Let us consider this more in detail. In the basis $(W^+$ $V^+)^T$ the mass 
square matrix is given by
\begin{equation}
\frac{g^2}{4}\,\left(\begin{array}{cc}
A+v^2_\rho/2& \sqrt{2}\,v_{\sigma_1}v_{\sigma_2}\\
 \sqrt{2}\,v_{\sigma_1}v_{\sigma_2}& A+v^2_\chi/2
\end{array}
\right),
\label{mixing}
\end{equation}
where $A=(v^2_\eta+v^2_{\sigma_2}+2v^2_{\sigma_1})/2$. We see from 
Eq.~(\ref{mixing}) that if $v_{\sigma_1}=0$ there is no mixing between $W^+$
and $V^+$. The mass eigenstates are given by $W^{+}_i={\cal U}_{ij}B^+_j$, 
where $i,j=1,2$ and $B^{+}_1=W^+$, $B^{+}_2=V^+$ and the orthogonal matrix 
is given by ($N$ is a normalization factor)
\begin{equation}
{\cal U}=\frac{1}{N}\,\left(
\begin{array}{cc}\frac{r-s-\sqrt{4b^2+(r-s)^2}}{2b} & 1 \\
\frac{r-s+\sqrt{4b^2+(r-s)^2}}{2b}  &1 
\end{array}
\right)\approx 
\frac{1}{\sqrt{s^2+r^2}}\left( \begin{array}{cc}
-s&b\\
b& s
\end{array}
\right),
\label{mu}
\end{equation} 
where $r=v^2_\rho/2$, $s=v^2_\chi/2$ and $b=\sqrt{2}\,
v_{\sigma_1}v_{\sigma_2}$.

This mixing may have interesting cosmological 
consequences since there are interactions 
like $\kappa (g^2/\sqrt2)\bar{l}_{L}\gamma^\mu \nu^cW^-_\mu$. 
Notice that its strength 
depends on the small parameter $\kappa\propto v_{\sigma_1}v_{\sigma_2}/
v^2_\chi$ and it cam be neglected in the usual processes. 
In fact, the model has three different mass scales since $v_{\sigma_1}\ll v_i
\ll v_\chi$ with $v_i=v_\eta,v_\rho,v_{\sigma_2}$~\cite{mdp}. 
It means that $W^{+}_1\approx W^+$, $W^{+}_2\approx V^+$ with
$M^2_W/M^2_Zc_W$ compatible with the experimental data if $v_{\sigma_1}
\leq3.89$ GeV~\cite{mdp}.
However, the mixing between $W^+$ and $V^+$
is interesting once there are new contributions to the majoron emission.
In fact, because of the $W^+W^-M^0$ vertex we have the neutrino transitions 
$(\nu_h)_L\to (\nu_l)_L M^0$; because of the vertex $V^+V^-M^0$ we have 
anti-neutrino transitions $\overline{(\nu_h)}_R
\to \overline{(\nu_l)}_R$. 
Both contributions could be negligible since they are proportional to 
$v_{\sigma_1}$. More interesting is that it is possible to have 
neutrino--anti-neutrino transitions like the decay 
$(\nu_h)_L\to \overline{(\nu_l)}_RM^0$ mediated by the mixing between $W^+$ 
and $V^+$ as it is shown in Fig.~1. This diagram is ultraviolet finite in the 
Feynman gauge and depends quadratically on a low energy scale that we have 
chosen, conservatively, as being the $m_\tau$ mass. 
The latter process implies a neutrino width which is, in a suitable 
approximation, dominated by the tau lepton contribution and is given by
\begin{equation}
\Gamma=\frac{1}{8\pi^5}\, G_F^4\vert {\cal K}_{\tau 3}\vert^2 
\vert{\cal K}_{\tau1}\vert^2 m^3_{\nu_h}m^4_\tau v^2_{\sigma_2}\;
\left(\frac{M_W}{M_V}\right)^4\;
\label{width}
\end{equation}
where we have neglected a logarithmic dependence on $m_{\nu_h}$ ($\nu_h$ can 
be $\nu_\tau$ or $\nu_\mu$ with $\nu_l=\nu_\mu, \nu_e$ in the first case and 
$\nu_l=\nu_e$ in the second one). 
With reasonable values for the masses in 
Eq.~(\ref{width}), that is $m_{\nu_h}\approx1$ MeV for the case of the 
tau-neutrino and $M_V\approx400$ GeV we can get a width of the order of 
$10^{30}$ MeV (up to the suppression of the mixing matrix ${\cal K}$). 
The age of the universe has a correspondent width of
$10^{39}$ MeV, thus it means that 
the decay can have a lifetime less than the age of the universe and could be 
of cosmological interest. 
From the cosmological point of view there are also the 
processes $\nu_h+\nu_h\to M^0\to\nu_l+\nu_l$ and $\nu_l+\nu_l\to M^0+M^0$ 
which occur at the tree level approximation. The cosmological effect of these 
processes are the same as in Ref.~\cite{ggn}.

If the parameters in this model are such that the majoron is irrelevant from 
the cosmological point of view, there is still the possibility 
that the majoron may be detected by its influence 
in neutrinoless double beta decay with majoron emission 
$nn\to e^-e^-M^0$ (denoted by $(\beta\beta)_{0\nu M}$). However it needs the 
majoron-neutrino couplings in the range 
$m_\nu/v_{\sigma_1}\sim 10^{-5}-10^{-3}$ in order to 
have the majoron emission experimentally observable~\cite{bsv}.  
Notice that in the present model the accompanying $0^+$ scalar, which is
by definition the lightest scalar $H^0_1$, may not be emitted in 
$(\beta\beta)_{0\nu M}$ if it is a heavy scalar or it is very suppressed
by the mixing factor. 

In our model both, the usual neutrinoless $(\beta\beta)_{0\nu }$ decay and 
also the decay $(\beta\beta)_{0\nu M}$, have new contributions.
If ${\cal F}$ is conserved in the scalar potential or $v_{\sigma_1}=0$
the mixing among singly charged scalars occurs with $\eta^-_1$ and $\rho^-$
and between $\eta^-_2$ and $\chi^-$. However if we allow ${\cal F}$ breaking
terms in the scalar potential or $v_{\sigma_1}\not=0$ there is a general
mixing among the scalar fields of the same charge. 
For instance, the trilinear term $f_2\chi^T S^\dagger\rho$ in the potential 
in Eq.~(\ref{potential}) implies 
the trilinear interaction $f_2h^-_2h^-_2\chi^{++}$ and since there is a
general mixing among all scalars of the same charge it means
that there are processes where 
the vector bosons are substituted by scalars since the vertex 
$h^+_2e^-\nu$ does exist and $h^+_2$ mixes with all the other singly charged 
scalars. There are also trilinear contributions arisen because of the vertices
$W^-V^-H^{++}_1$ and $h^-_2h^-_2H^{++}_1$ as in Refs.~\cite{sv2,ep}. There is 
also the vertex $h^-_2h^+_2M^0$ contributing to the $(\beta\beta)_{0\nu M}$. 
It seems that the analysis of both  $(\beta\beta)_{0\nu }$ and 
$(\beta\beta)_{0\nu M}$ decays is more complicated than that that were 
considered in Ref.~\cite{mdp,ps}.

There are also phenomenological constraints on majoron models coming from 
search in laboratory of flavor changing currents like 
$\mu\to e+M^0$~\cite{exp1} or in astrophysics through processes like 
$\gamma+e\to e+M^0$ which contributes to
the energy loss mechanism of stars~\cite{ggn}. However, in the present model 
the majoron couples only to neutrinos, for quarks and electrons the 
couplings arise only at the 1 loop level. Hence, all these processes do not
constraint the majoron couplings at all (at the lowest order). 
The interaction of the majoron with neutrinos is diagonal in flavor. 
The coupling between the majoron and the real scalar field $H^0_j$, of the 
form $M^0M^0H^0_j$, is 
\begin{equation}
i{\cal O}_{4j}(\lambda_{10}+\lambda_{11})\frac{v_{\sigma_1}}{2},
\label{3e}
\end{equation}
which is a small coupling. Note that since the majoron decouples from the 
other imaginary parts of the neutral scalars there are no trilinear couplings 
like $M^0A^0H^0_j$ where $A^0$ denotes a massive pseudoscalar, hence the model
does not have the phenomenological consequences in accelerator physics as 
the seesaw majoron model does~\cite{diaz}. 

Finally a remark. Here we have assumed that there is no spontaneous
CP violation. Hence, all vacuum expectation values are real. If we
allow complex VEV it has been shown that CP is violated
spontaneously~\cite{laplata1}. If this is the case we have a mixing among
all the scalars fields and also the majoron mixes with all the others CP even
and CP odd scalars.  



\acknowledgments 
This work was supported by Funda\c{c}\~ao de Amparo \`a Pesquisa
do Estado de S\~ao Paulo (FAPESP), Conselho Nacional de 
Ci\^encia e Tecnologia (CNPq) and by Programa de Apoio a
N\'ucleos de Excel\^encia (PRONEX). One of us (CP) would like to thank
Coordenadoria de Aperfei\c coamento de Pessoal de N\'\i vel Superior (CAPES) 
for financial support.

\appendix
\section{Constraint equations}
\label{sec:aa}
\begin{eqnarray}
&&t_\eta=\mu^2_1v_\eta+\lambda_1v^3_\eta+\frac{\lambda_4}{2}v^2_\rho v_\eta+
\frac{\lambda_5}{2}v^2_\chi v_\eta+\frac{\lambda_{12}}{2}
(v^2_{\sigma_1}+v^2_{\sigma_2})v_\eta-
\lambda_{17}v^2_{\sigma_2}v_\eta+\frac{\lambda_{19}}{2}v^2_{\sigma_1}v_\eta
\nonumber \\ &-& \frac{\lambda_{15}}{2\sqrt2}v^2_\chi v_{\sigma_2}
\frac{\lambda_{16}}{2\sqrt2}v^2_\rho v_{\sigma_2}+
\frac{f_1}{2\sqrt2}v_\rho v_\chi, \\ &&
\label{vin1}
t_\rho=\mu^2_2v_\rho+\lambda_2v^3_\rho+\frac{\lambda_4}{2}v^2_\eta v_\rho+
\frac{\lambda_6}{2}v^2_\chi v_\rho+\frac{\lambda_{14}}{2}(v^2_{\sigma_1}+
v^2_{\sigma_2}) v_\rho+\frac{\lambda_{16}}{\sqrt2}v_{\sigma_2}v_\eta v_\rho +
\frac{\lambda_{20}}{4}v^2_{\sigma_2} v_\rho\nonumber \\ &+&
\frac{f_1}{2\sqrt2}v_\eta v_\chi 
 +\frac{f_2}{4}v_{\sigma_2}v_\chi, \\ &&
\label{vin2}
t_\chi=\mu^2_3 v_\chi+\lambda_3v^2_\chi+\frac{\lambda_5}{2}v^2_\eta v_\chi+
\frac{\lambda_6}{2}v^2_\rho v_\chi+\frac{\lambda_{13}}{4}(v^2_{\sigma_1}+
v^2_{\sigma_2})v_\chi-
\frac{\lambda_{15}}{\sqrt2}v_{\sigma_2}v_\eta v_\chi +
\frac{\lambda_{18}}{4}v^2_{\sigma_2}v_\chi\nonumber \\ &+&
\frac{f_1}{2\sqrt2}v_\eta v_\rho+
\frac{f_2}{4} v_{\sigma_2}v_\rho, \\ &&
\label{vin3}
t_{\sigma_1}=\mu^2_4
v_{\sigma_1}+\lambda_{10}(v^2_{\sigma_1}+v^2_{\sigma_2})v_{\sigma_1}+
\lambda_{11}v^3_{\sigma_1}+\frac{\lambda_{12}}{2}
v^2_\eta v_{\sigma_1}+\frac{\lambda_{13}}{4} v^2_\chi v_{\sigma_1}+
\frac{\lambda_{14}}{2}v_\rho v_{\sigma_1}+
\frac{\lambda_{19}}{2}v^2_\eta v_{\sigma_1}, \\ &&
\label{vin4}
t_{\sigma_2}=\mu^2_4
v_{\sigma_2}+\lambda_{10}(v^2_{\sigma_2}+v^2_{\sigma_1})v_{\sigma_2}+
\frac{\lambda_{11}}{2}v^3_{\sigma_2}+\frac{\lambda_{12}}{2}v^2_\eta 
v_{\sigma_2}+\frac{\lambda_{13}}{2}v^2_\chi v_{\sigma_2}+
\frac{\lambda_{14}}{2}v^2_\rho v_{\sigma_2}
 - \lambda_{17}v^2_\eta v_{\sigma_2}\nonumber \\ &+&
\frac{\lambda_{18}}{4}v^2_\chi v_{\sigma_2} 
+\frac{\lambda_{20}}{4}v^2_\rho v_{\sigma_2}
-\frac{\lambda_{15}}{2\sqrt2} v^2_\chi v_\eta 
+\frac{\lambda_{16}}{2\sqrt2} v^2_\rho v_\eta 
+\frac{f_2}{4}v_\rho v_\chi. 
\label{vin5}
\end{eqnarray}




\vglue 0.01cm
\begin{figure}[ht]
\centering\leavevmode
\epsfxsize=\hsize
\epsfbox{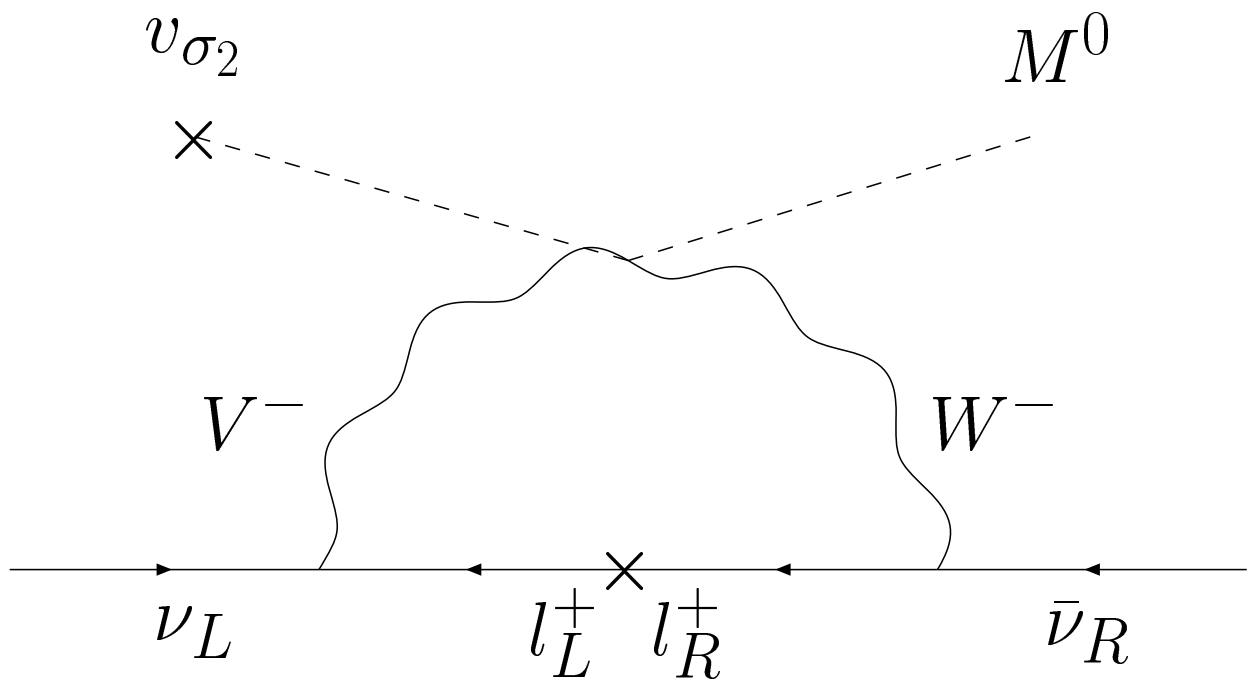}
\vglue -0.01cm
\caption{ One loop contribution to the process $\nu_L\to 
\bar{\nu}_R+M^0$.}
\label{fig1}
\end{figure}


\begin{references}
\bibitem{cmp1} Y. Chikashige, R. N. Mohapatra and
R. D. Peccei, Phys. Lett. {\bf98B}, 265 (1981).
\bibitem{gr} G. B. Gelmini and M. Roncadelli, Phys. Lett. {\bf99B}, 411 (1981).
\bibitem{pdg} C. Caso {\it et al.}, The European Physical Journal, {\bf C3},
1 (1998).
\bibitem{ggn} H. Georgi, S. L. Glashow and S. Nussinov, Nucl. Phys. 
{\bf B193}, 297 (1981).
\bibitem{concha} M. C. Gonzalez-Garcia and Y. Nir, Phys. Lett. {\bf232}, 383
(1989).
\bibitem{bs} S. Bertolini and A. Santamaria, Nucl. Phys. {\bf B310}, 714 
(1988).
\bibitem{km} H. Kikuchi and E. Ma, Phys. Lett. {\bf B335}, 444 (1994).
\bibitem{sv1} J. Schechter and J. W. F. Valle, Phys. Rev. D {\bf25}, 774 
(1982).
\bibitem{331} F. Pisano and V. Pleitez, Phys. Rev. D {\bf46}, 410 (1992);
R. Foot, O. F. Hern\'andez, F. Pisano and V. Pleitez, Phys. Rev. D {\bf47},
4158 (1993). See also P. Frampton, Phys. Rev. Lett. {\bf69}, 2889 (1992). 
\bibitem{pt} V. Pleitez and M. D. Tonasse, Phys. Rev.{\bf D48}, 5274 (1993).
\bibitem{tj} This case was considered briefly by M. B. Tully and G. C. Joshi, 
hep-ph/9810282 but no detail of the majoron was shown.
\bibitem{diaz} M. A. Diaz, M. A. Garcia-Jareno, D. A. Restrepo and J. F. W. 
Valle, Nucl. Phys. {\bf B527}, 44 (1998).
\bibitem{eu} See E. Kolb and M. Turner, {\sl The Early Universe}, 
(Addison-Wesley, New York, 1990) and references therein.
\bibitem{cmp2} Y. Chikashige, R. N. Mohapatra and R. D. Peccei, Phys. Rev. 
Lett. {\bf45}, 1926 (1981).
\bibitem{mdp} J. C. Montero, C. A. de S. Pires and  V. Pleitez, 
Preprint-IFT-P.020/99; hep-ph/9902448; to be published in Physical Review D.
\bibitem{bsv} Z. G. Berezhiani, A. Yu. Smirnov and J. W. F. Valle, Phys. Lett. 
{\bf B291}, 99 (1992).
\bibitem{sv2} J. Schechter and J. W. F. Valle, Phys. Rev. D {\bf 25}, 2951
(1982).
\bibitem{ep} C. O. Escobar and V. Pleitez, Phys. Rev. D {\bf 28}, 1166 (1983).
\bibitem{ps} F. Pisano and S. Shelly Sharma, Phys. Rev D {\bf57}, 5670 (1998).
\bibitem{exp1}  D. A. Bryman and E. T. H. Clifford, Phys. Rev. Lett. {\bf57}, 
2787 (1986).
\bibitem{laplata1} L. Epele, H. Fanchiotti, C. Garc\'\i a Canal, D. G\'omez 
Dumm, Phys. Lett. {\bf B343}, 291 (1995).
\end{references}
\end{document}